# Viability of Ultrasonic Sonochemical processing for nanostructures: Case study of Aluminum-crystal growth and Poly (vinylpyrrolidone)-graphitization

S.K. Padhi [a,†], M. Ghanashyam Krishna [a,b]

a. *School of Physics and Advanced Centre of Research in High Energy Materials, University of Hyderabad, Hyderabad 500046, India. †Email: **spadhee1@gmail.com***

b. *Centre for Advanced Studies in Electronics Science and Technology, School of Physics, University of Hyderabad, Prof C R Rao Road, Hyderabad 500046, Telangana, India.*

# Table of Contents

# Chapter-4

# Viability of Ultrasonic Sonochemical processing for nanostructures: Case study of Aluminum-crystal growth and Poly (vinylpyrrolidone)-graphitization

# Chapter-IV

# Abstract

The viability of ultrasonic sonochemistry is investigated in the context of air-stable metallic Al rich-PVP composite. The parameters investigated are; sono-(1) process intensification, (2) crystallization, (3) agglomeration, and (4) fragmentation, respectively. The conventional solvent of n-hexadecane is employed as the sonotrode generated pressure transmitting medium to carry out the above experiments. Two precursors, (1) Poly (vinylpyrrolidone) (PVP) and (2) Aluminum chloride ($AlCl_3$) are chosen to evaluate and demonstrate the viability of the ultrasonic induced processing. Temperature controlled investigations at RT and higher temperature help in achieving; (a) PVP-graphitization, and (b) Al-crystal growth phenomenon, respectively. The current experiments aid in helping to isolate and identify actual mechanistic happenings. The investigation, thus, has a fabrication protocol of shortened processing-duration, native amorphous oxide-free, metal-rich air stable product that leads to 10 g of composite product for fuel applications.

# Graphical Abstract

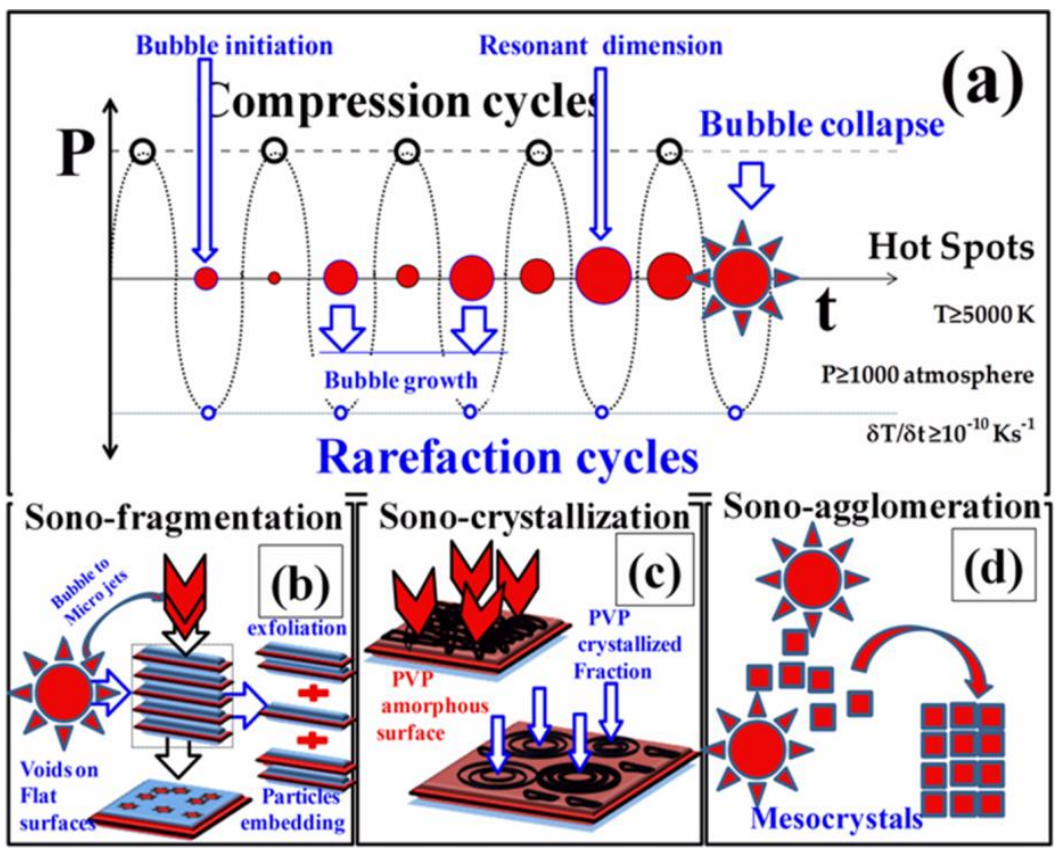

[Ultrasonic Sonochemical Viable for Nanoscience] Schematic presentation of: (a) Ultrasonic pressure waves leading to initiation of cavitation to impulsive collapse generating extreme conditions, (b) Sono-fragmentation (exfoliation, particle intercalation in 2D-materials), (c) Sono-crystallization in PVP, and (d) Sono-agglomeration of Al crystals building units generating large 2D mesocrystalline lumps, respectively.

## 4.1 Introduction

The widespread applicability of ultrasound under environmentally benign conditions delivering industrial scale product quality enrichment and production is fascinating. The areas of application include: food science and its associated technology (processing, preservation and extraction) development [54]–[60], water remediation [61]–[66], biomedical field [67]–[69], and also the process intensification of variety of processes [59], [70]–[73], etc respectively. It is a demonstrated fact that ultrasound-assisted protocol is more effective than that its corresponding conventional (physical, chemical, and biological) approach [74], [75]. In this context, the use of ultrasound in process intensification to deliver organometallic complexes (e.g., organo-lithium, -magnesium, and –aluminum, etc.) can be traced back as early as the 1950s, demonstrating its utility [76]. A few well known specific cases of metals activation by ultrasound leading to substantially shortened reaction duration (sonic acceleration) demonstrations of synthetically significant protocols are; (1) Zinc-induced Reformatsky reaction [77], (2) Copper-induced Ullmann couplings [78], and (3) Lithium-induced Barbier reaction [79], respectively. In addition to the specific cases, the principles of ultrasound-induced activation of metals and its use in accelerating (process intensification) organic synthesis are reported in terms of book chapters [80]–[84]. The point being ultrasound-induced shortened-in-time synthetic protocol development is an ongoing aspect. It is worth noting here that the ultrasound-induced cavitation and its cavitation impulsive collapse generated mechanical effects (like liquid microjets, turbulent mixing, shock waves, and acoustic streaming) are the dominant phenomena responsible for these synthetic process sono acceleration [85].

    Since its invention in 1927, sonocrystallization is another significant physio-chemical process of ultrasonic sonochemistry [86]–[89]. Investigations on possible mechanistic reasoning of sonocrystallization are actively ongoing. The question whether

sonocrystallization is ambient RT and mostly athermal shock wave-dominated phenomena needs exploration. In fact demonstration of the sonocrystallization phenomenon includes (1) aspirin as model for the molecular crystal [90], (2) organic molecules [91], (3) alkali halides as the ionic crystals [92]. The outputs of such studies suggest it is the direct particle and shock wave interaction is responsible for facilitating such phenomena. Still, a general acceptance of sonocrystallization out of these few individual case studies on crystallization and acoustic cavitation is not sufficient. This has also been suggested in a recent detailed review of possible mechanisms of sonocrystallization in solution [93], [94].

Sono-agglomeration as a result of a high-velocity inter-particle collision and subsequent fusion by melting delivered grain growth is also another significant physio-chemical attribute. Literature on some unusual sonochemical-assisted assemblies developed are: (1) graphene oxide (GO) and carbon nanotube (CNT) [95], (2) 2D-materials (graphene, $MoS_2$, h-BN etc) on flexible polymer substrates [96], (3) mesocrystals of $TiO_2$ and $BaTiO_3$ [97]–[100], and (4) silica spheres [101] etc. The case of metals sono-agglomeration during sonoprocess is extensively investigated by Suslick and co-authors et al [102]–[104]. Two particular outcomes of the metal sono-agglomeration studies are; if (1) particles collide head-on, it leads to agglomeration; otherwise if (2) the collision is at glancing angle leads to the removal of the inbuilt respective metals surface oxide layer by cracking and finally making the surface highly reactive. Thus, sonocrystallization leads to the generation of the crystalline nuclei while sono-agglomeration drives these generated nuclei to coalescence resulting in building unit and crystal growth.

Given these promising physio-chemical viables of ultrasonic sonochemistry, this chapter is an attempt to realize, demonstrate, and quantify the phenomenon like; (a) process intensification of chemical reaction, (b) sonocrystallization, (c) sono-aggregation. These phenomenon are investigated using precursors; (1) N-polyvinyl pyrrolidone (PVP), and (2) anhydrous $AlCl_3$ in conventional hexadecane solvent respectively, as case studies. The PVP polymer is used to adjudge sonocrystallization phenomena without bulk solution heating at ambient laboratory conditions. This judgment is to isolate whether it is an athermal shock wave generated shear/pressure linked or temperature linked process. In contrast, metallic Al is used to understand the crystal growth aspect employing "ultrasonic-assisted process intensification activity" based on conventional solution-phase chemical Al precursor reduction process to deliver Al nanoparticles. As stated, sonoprocess generated Al nanoparticles of uniform dimension and surface oxide-free are incorporated into the sonocrystallized PVP matrix. The motivation of this chapter is employing sono-process for achieving the fabrication PVP (P), graphitic carbon (GC), and Al (M) incorporated air-stable composite. Further is to examine the loss of metallic content after year-long storage, which is essential for fuel application.

In this context, generic protocols to fabricate oxide-free Al nanocrystals involve either; (1) direct solution-phase reduction of Al precursor reduction leading to Al crystal growth or via a (2) alane-precursor based thermal decomposition schemes [306]–[308]. In these schemes for safe laboratory handling, surface passivation of the Al nanostructured product is achieved either by (a) controlled air exposure (help in developing thin amorphous $Al_2O_3$ outer shell) or (2) an appropriate polymer surface coating respectively. Most importantly, these protocols run over several long hours to complete. It is important to note that, besides sono-chemical, attempts to superimpose with electric and microwave field stimulation on many conventional approaches for reaction process intensification is also reported [309], [310]. Significantly, the introduction of sonochemical stimulation to the organic alane-precursor based thermal decomposition (protocol-2) reaction scales down, remarkably, to just several minutes [311]. Although the use of inorganic Al precursor also attempted in sono-electrochemical, electrochemical template deposition and polymer stabilization, the process still runs over several hours resulting in non-uniform larger particle size, scale-up limitations, and energy content inefficiencies [312]–[315]. The inorganic case (protocol-1) is the most utilized conventional case left to be investigated employing heterogeneous sonochemistry as standalone stimulation for process intensification studies.

## 4.2 Materials and Methods

Chemicals and precursors used in this chapter are of Analytical Reagent (AR) grade. Chemicals received from the different vendors are used without any further purification. Aluminum chloride ($AlCl_3$, Reagent plus (R), 99%), Lithium Aluminum Hydride ($LiAlH_4$, pellets, Reagent grade, 95%), Poly (vinylpyrrolidone) (PVP, molecular weight 10,000), n-Hexadecane ($CH_3(CH_2)_{14}CH_3$, anhydrous, 99%), and UHP Argon are used. Glassware related accessories cleaned by standard laboratory procedures, and the nitrogen glove box is used to handle moisture-sensitive chemicals.

Sonochemical processing is carried out with Sonics VCX-750 watt ultrasonic processor. Sonochemical reaction vessel (40-250 mL processing capacity, three 14/20 side necks, glass chamber height 62 mm), with the adapter (Part number: 830-0014) is screwed into the special long full-wave solid probe (Titanium alloy Ti-6Al-4V, 13 mm tip, 245 mm long) at the nodal point. The glass sonochemical reaction vessel slides on the adapter and is fixed in a place as required by the bushing which is screwed into the reaction vessel, with an O-ring compress. The reaction vessel movement on the adapter facilitates the probe portion extension out of the adapter required to be immersed into the sample. A continuous mode of operation for 2 hrs (process control from 1s to a maximum of 10 hrs) processing is carried out with ice water (20 °C) circulation based on the requirement. The UHP argon bubbling at 30 bubbles /minute is also maintained during sonoprocessing. The snapshots of the sonochemical reaction vessel with a sonotrode arrangement are shown in figs. 4 1.

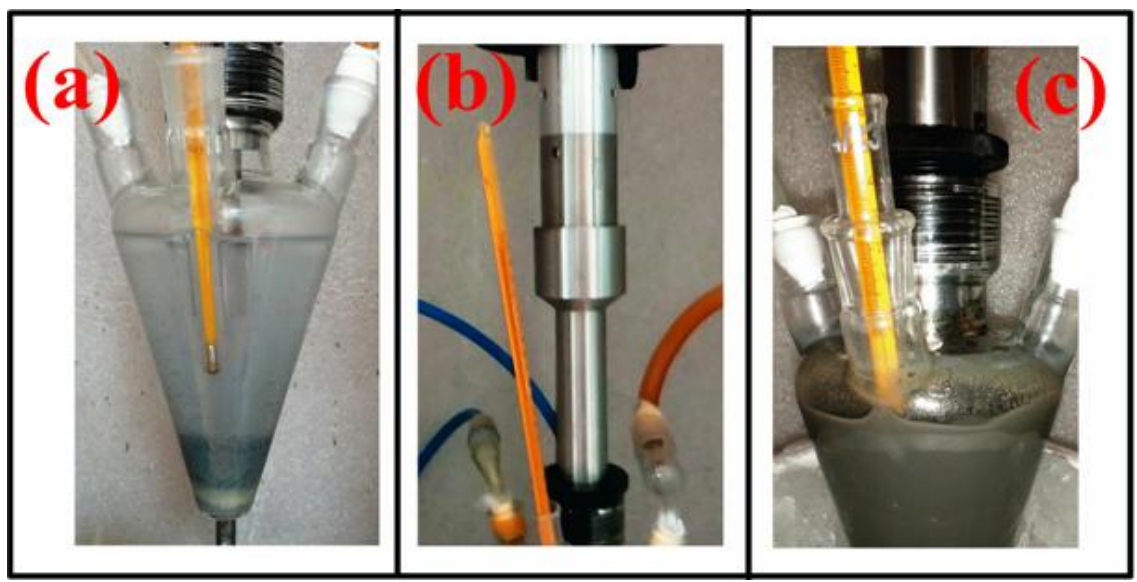

Fig.4 1 [*Sonochemical Reaction Vessel*]: Photographic snapshots (a) precursor before ultrasonication, (b) long full-wave solid probe fixed onto the adapter (c) after sonication respectively.

## 4.3 Results and Discussion

### 4.3.1 PVP TEM Investigations

#### 4.3.1.1 PVP Pristine

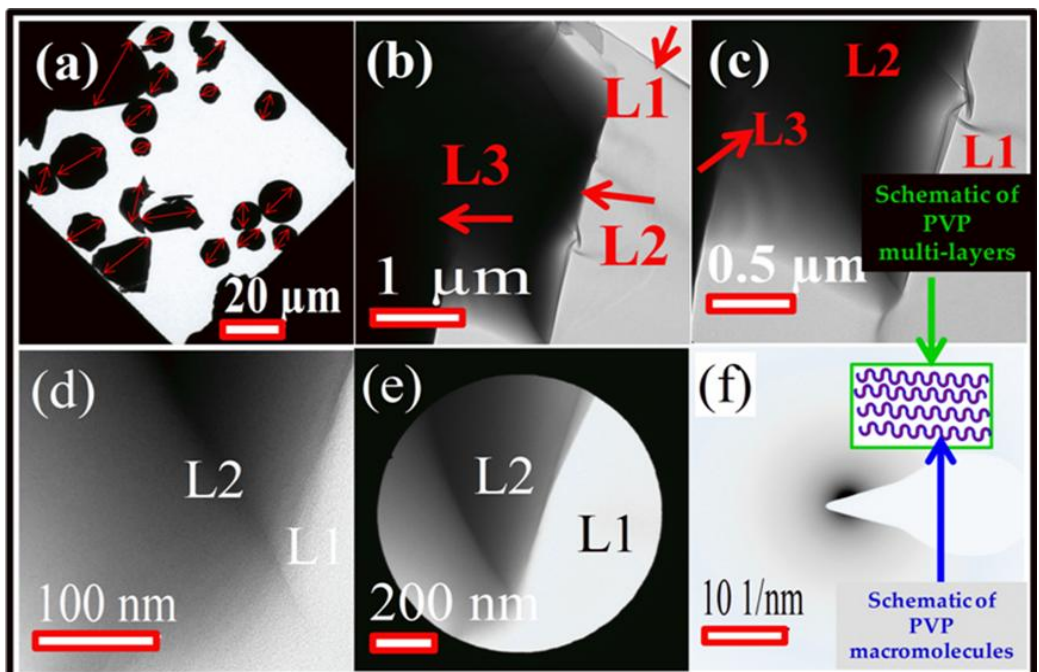

Fig.4 2 [*Pristine PVP TEM observations*]: TEM BF (a) lumps, (b)-(e) PVP layers, and TEM SAED aperture, (f) characteristic halo disc pattern overlaid with PVP as multilayered shell schematic as inset respectively.

PVP powder spread onto a TEM grid inside a nitrogen glove box is transferred to the TEM sample holder immediately and is imaged. Irregular shape μ-size bulky aggregates of PVP macromolecules bulky aggregates confined to one of the TEM grid square-mesh is shown in fig. 4 2 (a). Subsequent sequential increased magnification TEM BF images are recorded and are shown in figs. 4 2 (b)-(d). Individual lumps (see fig. 4 2 (a)) edge portions are imaged and depict layered morphology having smooth (no crystallized or foreign entities as an embedded fraction) surface microstructure. The increasing dark contrast (i.e., increased thickness) is a result of PVP macromolecules' layered aggregation in sequential fashion when observed from any of the bulky lump edges to the center. The TEM-SAED recorded from these layers has the characteristic halo-disc shape of amorphous materials. One such localized region with SAED aperture and obtained SAED pattern are shown in figs. 4 2 (e), and (f), respectively. The PVP material is stable under step-2 TEM e-beam investigation used for probing, as evident from lack of changes to these layers surface microstructures in the present illustrations. Based on the current TEM-BF study (figs. 4 2 (b)-(e)) and literature, a representative schematic of PVP lumps concurrent with the observations is overlaid on fig. 4 2 (f) as inset [316].

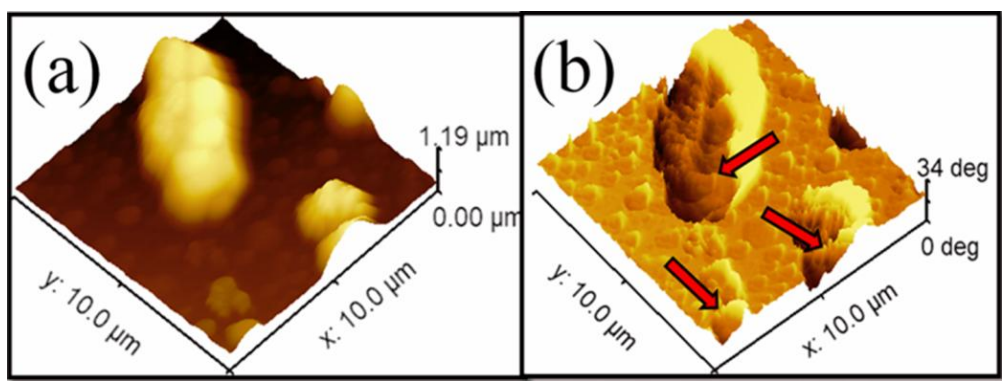

Fig.4 3 [Pristine PVP AFM observations]: DFM operation for obtaining; (a) 3D-topography, and (b) corresponding phase contrast image (arrows indicate lump wall) respectively.

Ethanol solvent dispersed PVP spin-coated on to a silicon substrate is imaged by employing the non-contact dynamic force microscopy (DFM) mode in an AFM. PVP lumps 3D view is acquired to support TEM 2D observations depicting no contrast. A larger PVP globule is chosen, and its acquired 3D topography is shown in fig. 4 3 (a). Acquired globule represents one TEM lump and is about micron thick, thereby non-transparent to TEM e-beam, hence is of darker contrast. Many micron-sized smaller spherical aggregates constituting this lump can be seen in topography, but are recorded with better contrast for differentiation in the phase image shown in fig. 4 3 (b) [317]–[319]. The existing individual aggregate walls are of 100-400 nm thick and are marked on the corresponding phase-contrast microscopy image with single-headed arrows (see fig. 4 3 (b)).

**4.3.1.2 PVP Sonicated**

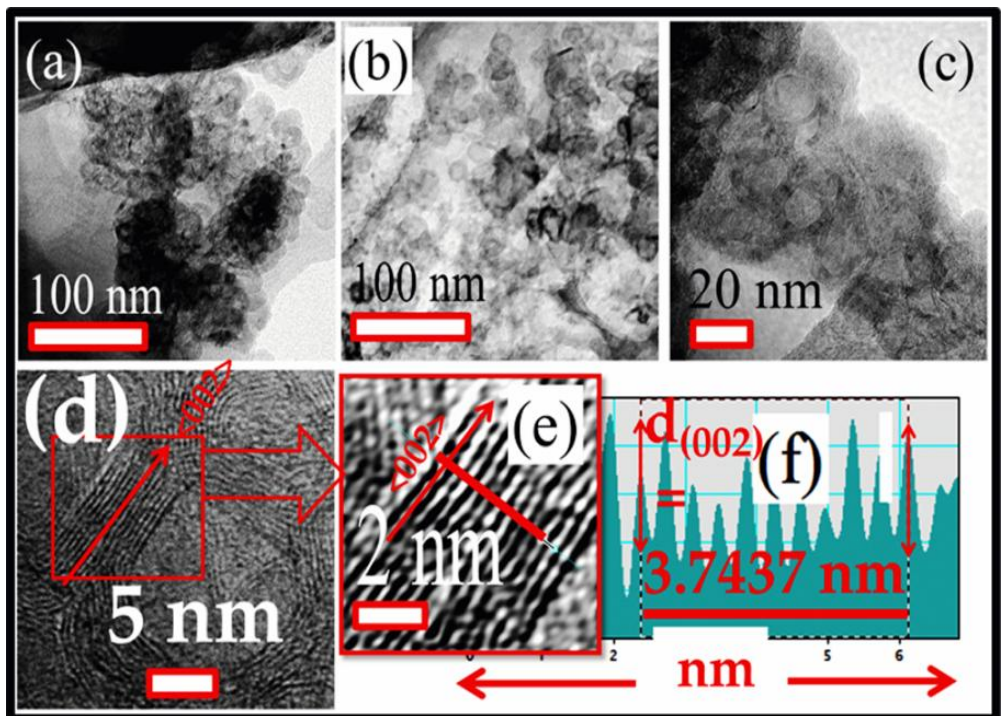

Fig.4 4 [*PVP sonicated TEM observations*]: TEM BF (a) lumps, (b)-(e) PVP layers, and TEM SAED (f) characteristic halo disc pattern respectively.

Reports of cavitational reactors delivered process intensification had many demonstrations [320]–[328]. In this context, 20 kHz ultrasound pressure wave's irradiation generated mechanochemical alterations to PVP polymer are investigated first. For this purpose, a 2 hrs long (previously optimized) ultrasound irradiation processed PVP polymer product transferred onto TEM grids are imaged. Sonochemical vessel of 250 mL capacity with 1.08 g of PVP at its bottom is ultrasonic irradiated (Sonics VCX 750W, 13 mm solid ultrasonic horn is used at 50 % amplitude) through hexadecane solvent as pressure wave transmitting medium. Out of many, the specific effects of ultrasonic irradiation generated signatures of importance specific to the current study observations are shown in figs. 4 4 (a)-(c) as TEM BF images. The two notable PVP polymer surface observed attributes are; (1) surface rupture, and (2) evolved randomly distributed crystalline features presence respectively. The first aspect is mechanical, a physical activity termed as sonofragmentation [329]–[336]. While the second feature highlights ultrasound application in solution mediated materials crystallization (otherwise known as sonocrystallization) processes, respectively [337]–[340].

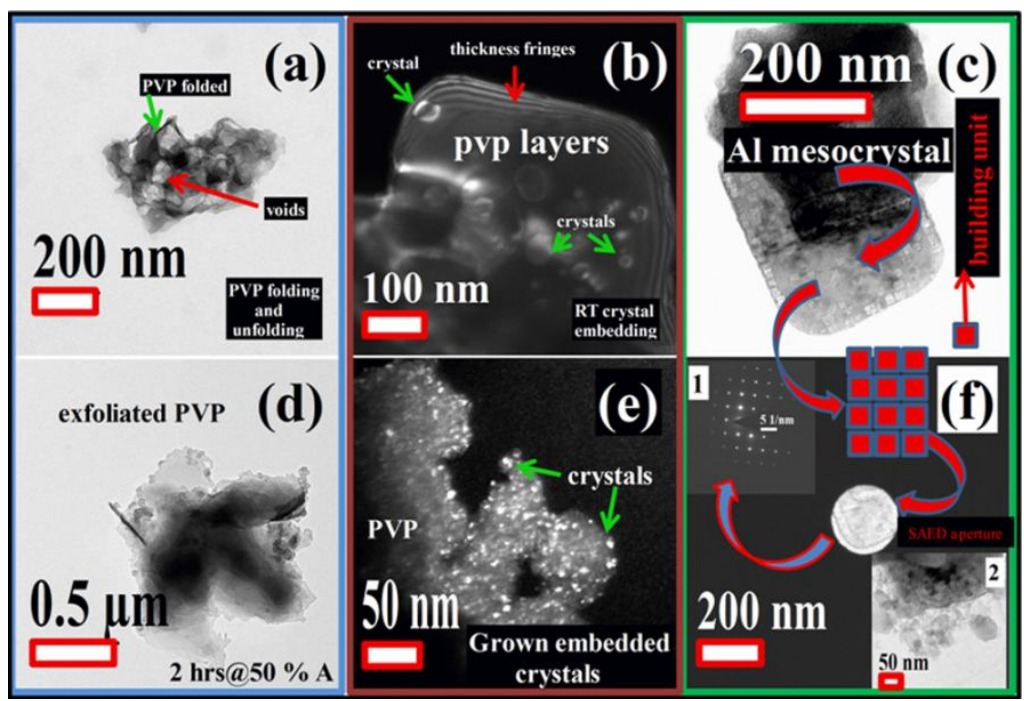

Fig.4 5 [*Sonochemical Mechanochemical deliverables*]: (a)-(d) exfoliation, (b)-(e) growth and embedding, and (c)-(f) aggregation respectively.

The major sonochemical attributes encountered are schematically presented as shown in figs. 4 5 (b)-(d). To achieve these, the impulsive bubble collapse impetus driven ultrasonic mechanochemistry is depicted in fig. 4 6 (a). TEM BF images, of ultrasonic irradiation hexadecane solvent medium processed products, acquired justify these occurrences are displayed in figs. 4 5 (a)-(f) respectively.

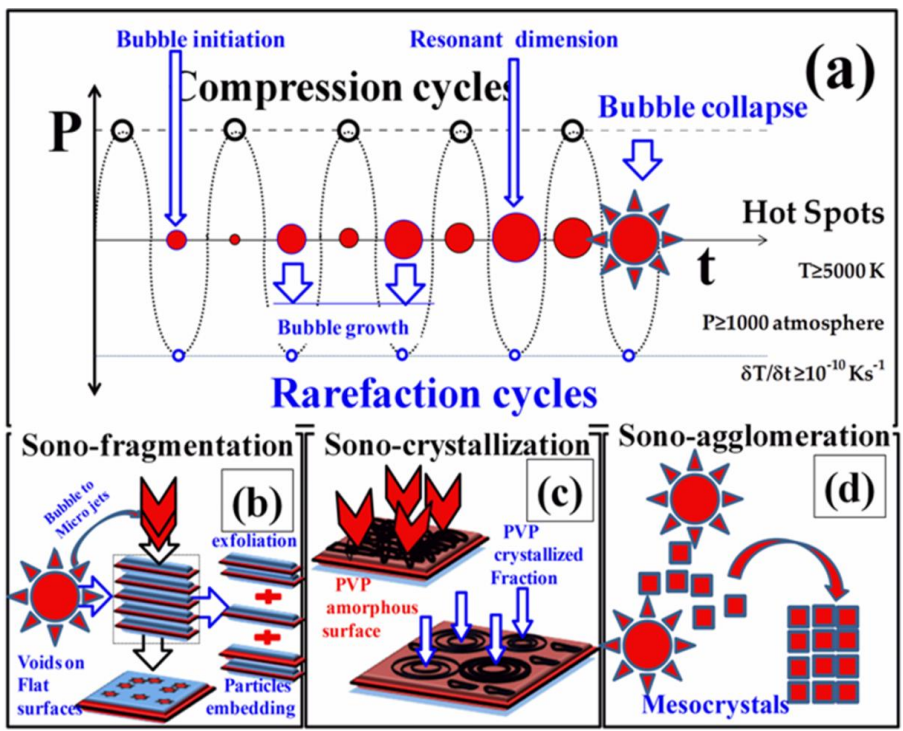

Fig.4 6 [*Sonochemical Mechanochemical deliverables*]: (a)-(d) exfoliation, (b)-(e) growth and embedding, and (c)-(f) aggregation respectively.

### 4.3.1.3 DFM probing PVP Sono-fragmentation

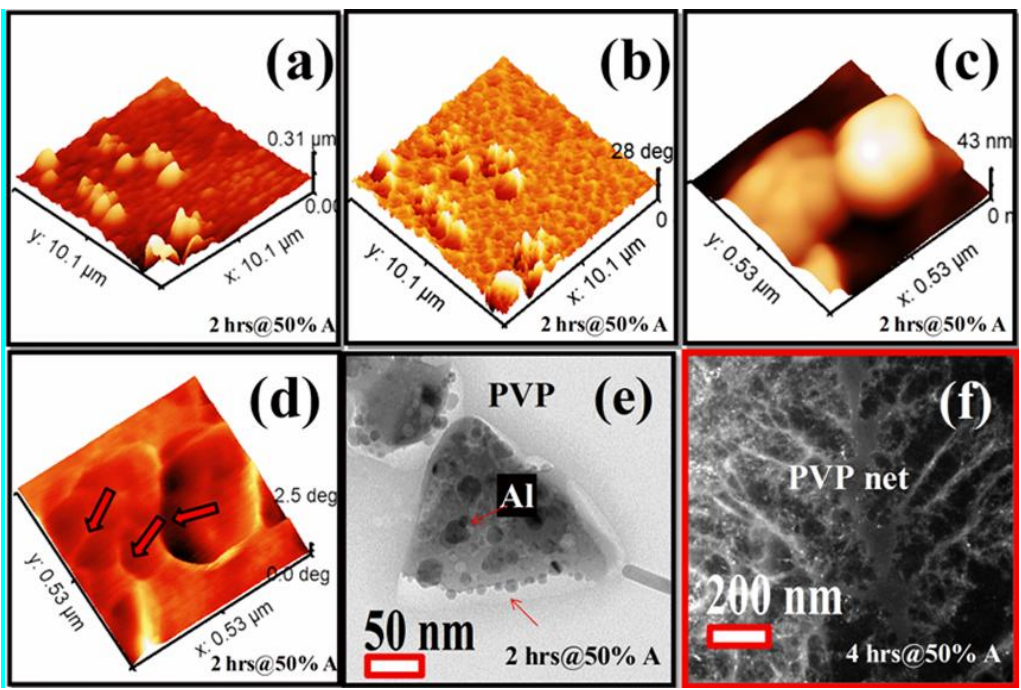

Fig.4 7 [*Sono-fragmentation*]: (a)-(d) DFM mode observation of PVP in topography and phase, (e) growth and embedding of nano-Al, and (f) PVP network after extended sonication respectively.

A detailed guideline for liquid-phase exfoliation (LPE) employing ultrasonication and its slightly modified, adapted techniques for 2D-layered materials published elsewhere is followed [341]–[348]. Three notable reasons delivering LPE identified are; (1) cavitational bubble collapse leading to stemming generated mechanical energy in the form of compressive/tensile stress wave in an unbalanced manner to overturn the inbuilt layers attraction, resulting exfoliation, (2) shock waves breaking bulk into thin flakes, (3) cutting of flakes due to frictional force resulting from high strain rates up to $10^9$ $s^{-1}$, and (4) combination of all these processes acting simultaneously respectively. In the present case, the fragmentation of PVP layers is achieved in hexadecane (Sonics VCX 750W, 13 mm solid ultrasonic horn is used at 50 % amplitude) ultrasonic irradiated for 2 and 4 hrs respectively. The non-contact DFM mode observation in both topography and phase shown in figs. 4 7 (a)-(d), imply thickness almost approachable to 50 nm indicating flat 2D nanostructured layers. One such layer having nano-Al embedded in it is shown in fig. 4 7 (e). Likewise, PVP sonicated for extended 4 hrs becomes network like and is, hence, not appropriate for nano-Al surface stabilization. This extracted product examined in TEM is observed to have around 80-90 nm Al particles wrapped in GC network. Also, the development of the amorphous-$Al_2O_3$ layer is seen to be developed after storing in laboratory environment for a week.

### 4.3.1.4 TEM probing PVP Sono-crystallization

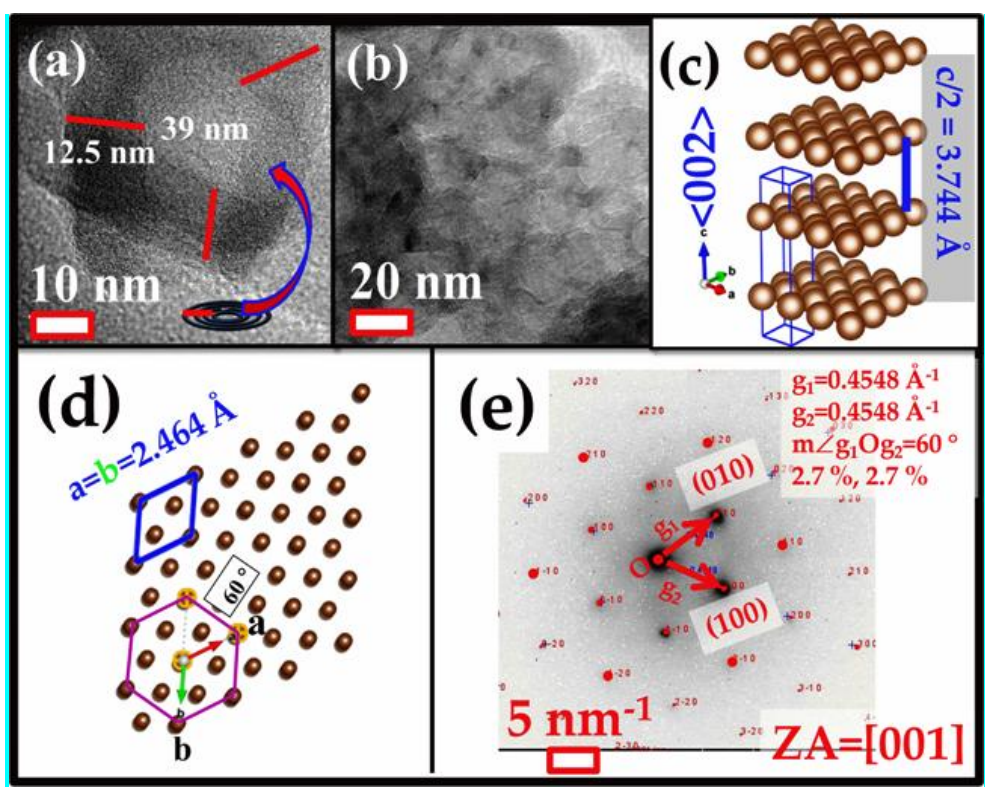

Fig.4 8 [*Sono-crystallization*]: (a) PVP surface initiation of onion like features, (b) densely populated such features, (c)-(e) microstructural evaluation respectively.

The use of ultrasound in delivering crystallization in pharmaceutical had widespread demonstrations, but the physical mechanism underlying this process physical happenings is still under exploration [333], [339], [349]. Sonocrystallization of poly-3-hexylthiophene (P3HT) chains to nanofibers by the application of the ultrasonic field is proposed based on nucleation and growth aspects. This is a consequence of ultrasound assist in delivering sufficient mechanical energy to overcome the local energy barrier to trigger a small crystalline nuclei nucleation [337]. The evolved crystalline nuclei act as the seed for the subsequent growth of large nanofibers. In this context, consistent with many previous reports, experimental validation highlighting PVP crystallization to graphitic carbon (GC) is shown in figs. 4 8 (a)-(b). Initiation of onion-like stripes after 1 h (see fig. 4 8 (a)) and filling of such stripes all over the PVP surface (see fig. 4 8 (b)) after 2 hrs of ultrasonication in hexadecane is observed. The TEM microstructural data from these generated structures locally in HR-TEM (see figs. 4 2 (d)-(f)) and SAED (fig. 4 8 (e)) mode confirms PVP crystallization. The microstructural data extraction and schematic presentation of the same shown in figs. 4 8 (c)-(e), indicates hexagonal GC along with c-axis tensile strained in comparison with that of the standard ICDD PDF-2: 89-7213 file.

**4.3.1.4 TEM probing Aluminum Sono-agglomeration**

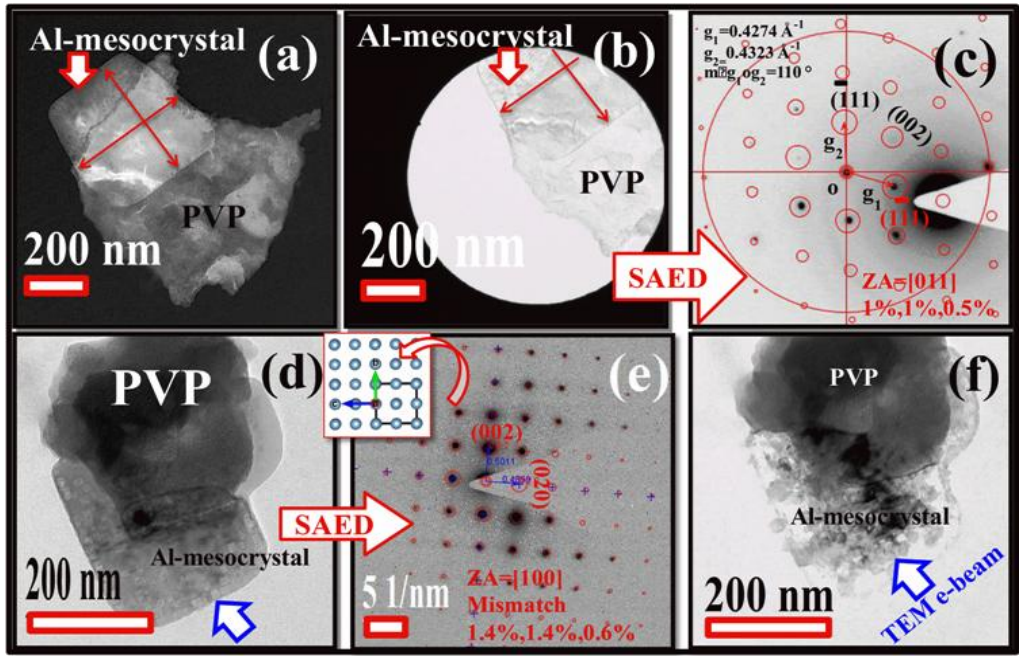

Fig.4 9 [*Sono-agglomeration*]: (a)-(d) Al mesocrystal, (b), (c), (e) TEM-SAED characterization, and (f) TEM e-beam de-agglomeration respectively.

Ultrasonic de-agglomeration is a frequently observed event, but literature on materials agglomeration during sonochemical processing is not rare. Ultrasonic's during sonoprocessing in generating agglomeration of metallic particles investigated by Suslick and group et al. had two interesting outcomes [350]–[355]. They are; if (1) particles collide head-on, the result is agglomeration, otherwise; (2) glancing angle collisions lead to surface oxide layer cracking and thereby its loss, respectively. In case of Al nanoparticles the surface oxide layer is a hindrance for its use as fuel; hence its growth is favorably inhibited (stated outcome 2) during sonoprocessing bringing about a positive development. The mesocrystalline Al formulations are shown in TEM BF/DF images in figs. 4 9 (a) and (d) are the implication of presented outcome 1. Likewise, TEM-SAED acquired, as shown in figs. 4 9 (c) and (e) is that of the Al structural phase. The identified zone axis from the experimental SAED implies a lattice mismatch of less than 2 % between that of the standard ICDD PDF: 04-0787 and experimental obtained

TEM-SAED pattern. Besides TEM-SAED, the investigation of Al particle surface (i.e., HRTEM mode) for the presence of surface oxide is attempted. However, HRTEM surface oxide isolation remained unsuccessful in isolating surface oxide validates the outcome 2 presented. The Al mesocrystal has shown in fig. 4 9 (d) just exposed to HRTEM mode e-beam exposure (E4I5M) initiates the disintegration of the mesocrystalline formulation of cubical building unit (see drawn schematic in fig. 4 3 (f)). The disintegrated Al mesocrystal after 5 minutes of step-4 HRTEM mode exposure is shown in fig. 4 9 (f). This implies Al cubical building units are loosely agglomerated (facile disintegration under TEM e-beam) but in a periodic coherent order to behave as a whole single crystalline block. Similar to present observation of Al mesocrystal formation under ultrasonic irradiation, case studies of materials orderly arrangement achieved in materials during sonoprocessing are; (1) $BaTiO_3$ mesocrystals, (2) layered arrangement of $CaCO_3$, and (3) $TiO_2$, etc [356]–[360].

Although the ultrasonic irradiation-induced inter-particle collision is the leading attribute contributing to sono-agglomeration, another essential contributor that needs mention linked to the solvent physical attribute (i.e., surface tension, viscosity, and vapor pressure, etc) used in sonolysis process. In brief preferred solvents having low viscosity, low surface tension, and less vapor pressure are the most preferred [361]. The list of conventional solvents mostly employed for sonochemical processing is; hexane, hexadecane, pentane, dichloromethane, etc [362], [363]. Also, in the case of polar (methanol, ethyl alcohol) vs non-polar solvent (diethyl ether, hexadecane) solvent effect during sonoprocessing for fabricating µ-CuO agglomerates; highlights non-polar solvents acts effectively [364]. The current Al mesocrystals extraction is done out of the hexadecane solvent fabricate Al-rich compositions with Al (M)/ PVP (P) ratio higher than 1:1 ratio. This product develops surface oxide after storage for a week in laboratory conditions, hence not useful for fuel applications.

## 4.4 Al Characterization

### 4.4.1 Nanostructured Al Stabilization

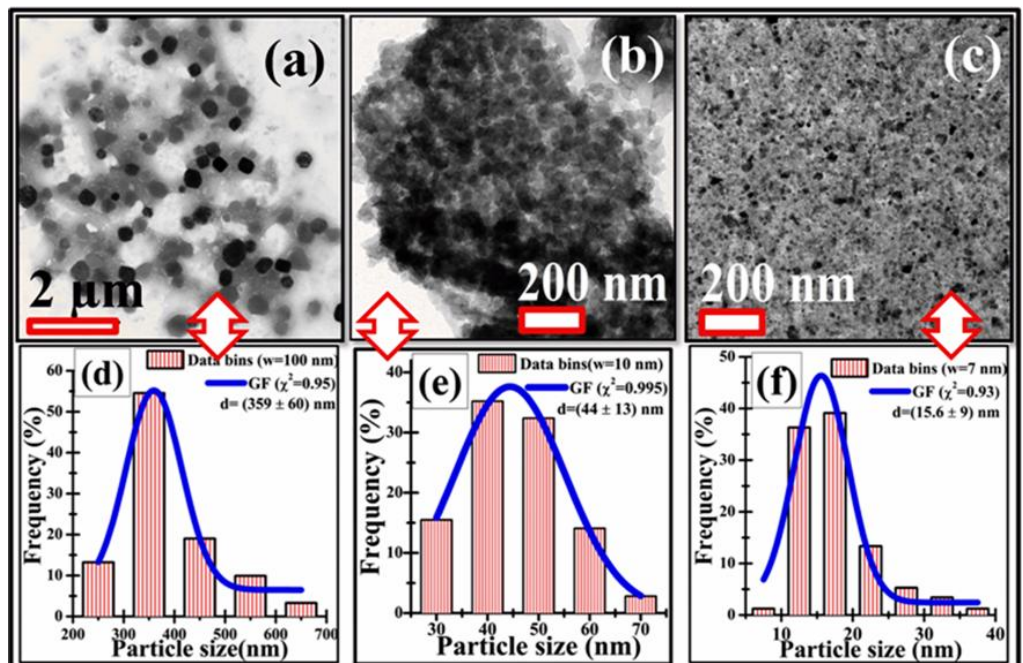

Fig.4 10 [*Nano-Al PVP surface stabilization*]: (a) 4:1 (b) 2:1, (c) 1:1 Al (M)/ PVP (P) compositions. (d), (e), (f) are the corresponding nano-Al particle size, respectively.

Synthesizing oxide-free Al nanoparticles stabilized in the PVP matrix, in gram quantities, for fuel application is the key objective. In doing so, the sonication induced "process intensification" activity is to be evaluated. Therefore one of the previously optimized Al chemical synthetic protocols is considered for experimentation. But an additional ultrasonic probe introduced to achieve "process intensification". There is published literature illustrating the specific chemical synthetic protocol to be replicated [365]–[367]. In order to have assertive quantification of the "process intensification" in a quantitative term a physical variable namely "degree of crystallinity (DOC)" linked to the crystalline Al diffracting volume fraction is chosen [368]–[370]. It is the integrated intensity of the crystalline Al diffracting component to that of the total integrated intensity of both the crystalline Al and amorphous PVP fractions. The process followed is to estimate DOC of the products is to employ Rietveld whole-pattern fitting method. Bruker AXS TOPAS (Total Pattern Analysis Solution) Version 5 program is used [371]–[373]. For analysis, the XRD broad signal from the amorphous phase is fitted with a split pseudo-Voigt (SPV) function. The peak position, the area, the left, and right FWHM, and the Lorentz fraction for the left and right SPV profiles are refined. The area under the curve of the SPV function is used as an effective scale factor for the amorphous phase.

The reflection profiles of crystalline phases are fitted with profile generated by fundamental parameter approach (FPA), most suited for diffractometer using Bragg-Brentano geometry [374]. The background intensity is modeled by Chebychev polynomial of $5$th order with 1/X background checked implemented in TOPAS. The implementation of this is subsequently done, but the synthetic chemical protocol to deliver Al rich fractions is attempted first. Those three Al (M)/ PVP (P) compositions to having Al theoretical DOC (T) of 80, 66, and 50 % are synthesized. Out of these, the 1:1 Al-PVP composite having DOC (T) =50 % has the smallest average Al particle size of (15.6±9) nm. Thereby this composite is the material of choice for subsequent further studies. The details of particle size distributions of these three Al-rich composites counted out of their corresponding TEM-BF images are shown in figs. 4 10 (a)-(f). It is significant to note that using intensified ultrasound-assisted approach to deliver nanostructured Al; (1) [bottom up chemical processing employing Al precursor] require 30 minutes of processing time [375], whereas in (2) [top down processing employing Al foil] takes almost 36 hrs [376]. Thus, in the ongoing experimentation, the ultrasonic "process intensification" brings down the sole chemical processing protocol from 24 hrs to just 2 hrs, based on DOC=50 % quantification for 1:1 Al-PVP composite.

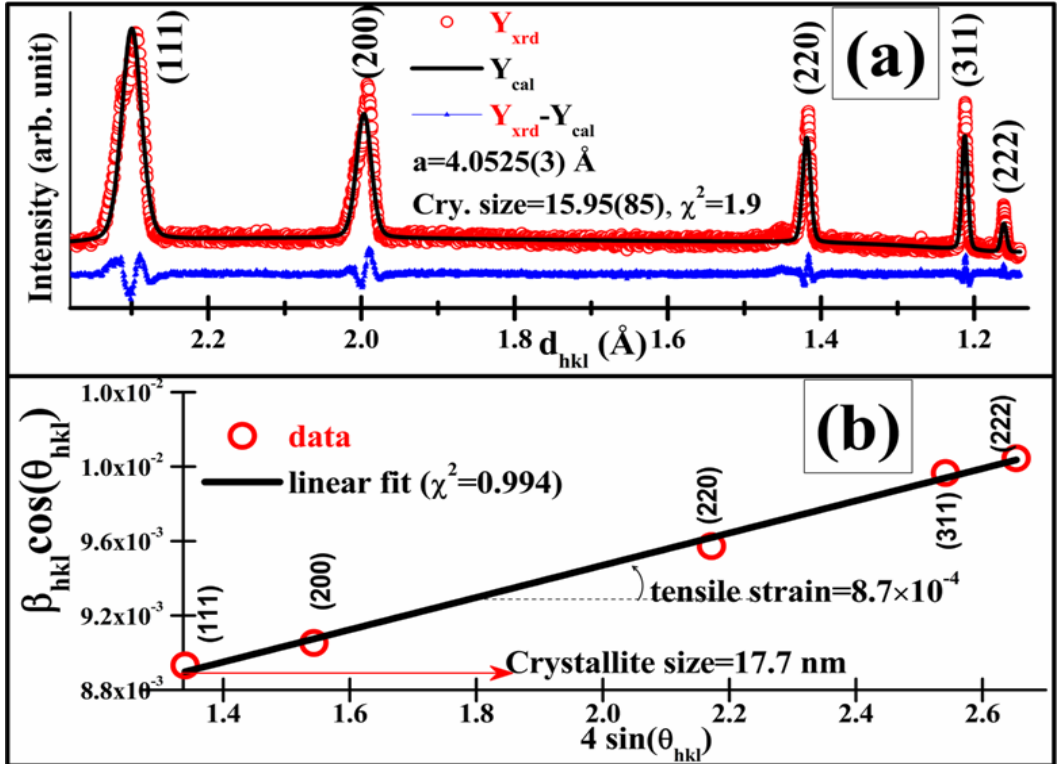

Fig.4 11 [*XRD structural analysis of nano-Al*]: (a) whole pattern profile fit, (b) WH plot respectively.

The lattice constant and phase purity of the embedded crystalline Al particles in the PVP matrix is estimated from the XRD data. The Al XRD data with profile fitting for lattice parameter extraction is plotted in fig. 4 11. The profile fitting refinement is terminated after reaching acceptable values of standard agreement triplets (weighted profile R factor ($R_{wp}$ %), expected R factor ($R_{exp}$ %), R-pattern ($R_P$ %), and goodness of fit index ($\chi^2$), with $\chi^2$=1 representing an exact refinement. The agreement triplets reached are 6.22, 11.13, and 8.60, with $\chi^2$=1.9, respectively. The obtained final profile fit and difference pattern are shown as $Y_{cal}$ and $Y_{diff} = Y_{xrd} - Y_{cal}$ in fig. 4 11 (a). The refined face-centered cubic (FCC) unit cell is tensile strained with a=4.052 Å and is higher than 4.049 Å representing standard ICDD PDF: 04-0787 file. The Williamson-Hall (WH) plot ($\beta Cos(\theta)$ Vs. $4Sin(\theta)$) taken from the (111), (200), (220), (311), and (222) miller indexed lattice planes is shown in fig. 4 11 (b). The slope of the fitted line is positive, providing a direct indication of the tensile strain state of the Al phase as evaluated by the profile fitting computation. No crystalline or amorphous characteristic of the oxide phase is observed, indicating phase purity of synthesized nano-Al.

### 4.4.2 Sonocrystallization of PVP at RT

The sono-mechanochemical driven PVP graphitization (i.e., sonocrystallization) process is investigated by using the bulk powder-XRD method. The analysis of such bulk XRD data is a reaffirmation and validation of the presented TEM localized microstructural graphitization. The specific outcomes being; (1) graphitized PVP fraction quantification, (2) graphitized carbon structural parameters evaluation, and (3) developed structural phase identification, respectively. A set of the sonochemical designed composites of x wt% PVP/y wt% Al (denoted as xPVP-yAl; where x/y=1/1, 2/1, and 4/1) products are processed. The XRD pattern of RT sonicated 3PVP-2Al composite (denoted as RTSC/PVP-Al) concurrent to the present discussion is plotted along with the parent-PVP in fig. 4 12.

The amorphous parent-PVP has characteristic broad humps at 2θ=11.6 and 20.2 °, respectively [377]–[380]. The broad hump at 2θ=20.2 ° develops to a sharpened peak implying PVP crystallization, along with its simultaneous structural phase evolution to graphitic carbon (GC) form. This process of crystallization and subsequent GC phase formation is achieved by probe sonication at RT in solution-phase chemical processing of the RTSC/PVP-Al composite product. The mechanistics of the ultrasonic pressure waves devised crystallization is similar to that observed under laser or electron beam [381]–[388]. The XRD pattern of RTSC/PVP-Al composite also highlights the process of; (1) intercalation, (2) growth, and (3) stabilization of metallic Al nanoparticulate phase achieved in the designed crystallized matrix of PVP and GC respectively.

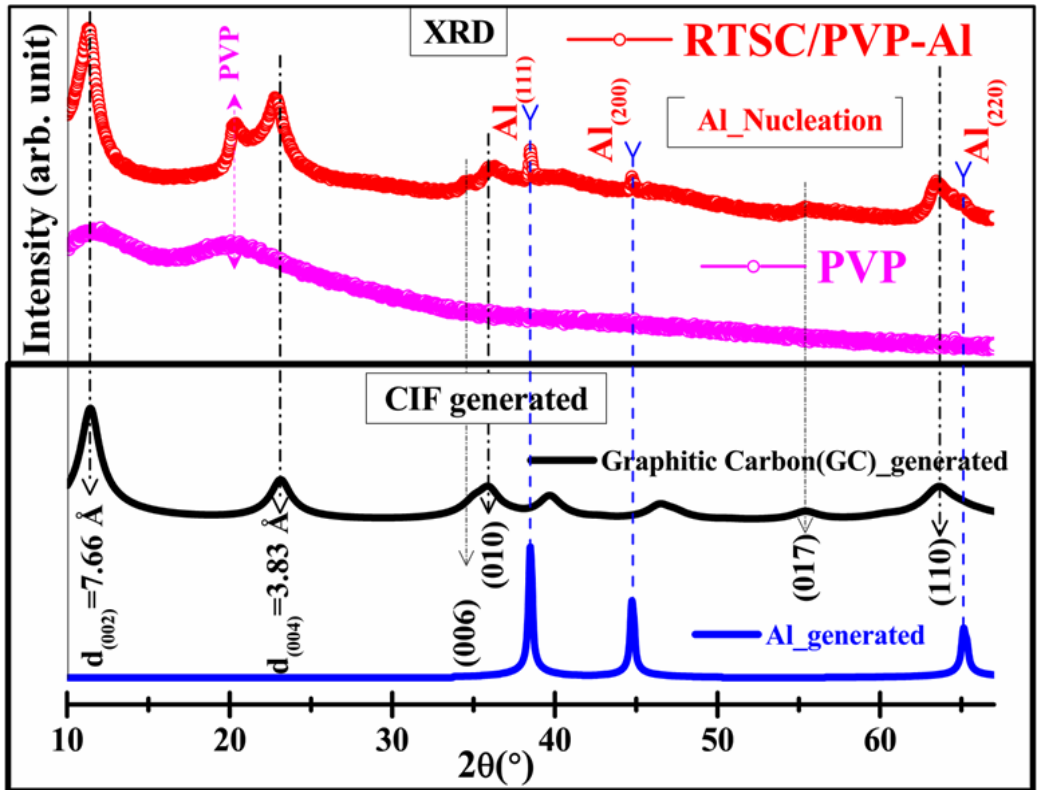

Fig.4 12 [*Sonocrystallization of PVP at RT*]: Obtained RTSC/3PVP-2Al composite XRD pattern plotted with crystal structure generated patterns below for developed peak phase identification.

In the designed RTSC/3PVP-2Al composite product, the PVP crystallized GC fraction structural phase identification is made employing Match! - program [389]. The fabricated GC structural phase has a match (i.e., the identified file is of the highest figure of merit) with that of the reference ICDD PDF-2: 89-7213 file with expanded C-axis. To reaffirm this further, the crystallographic information file (CIF) generated XRD patterns are included in the plot as Graphitic Carbon (GC) _generated in fig. 4 12. A perfect match between GC_ generated with that of the GC phase of the fabricated RTSC/3PVP-2Al composite product is elucidated for observation. Also, in continuation of the earlier discussions, the reference ICDD PDF-2 file: 04-0787 identified in the previous sections, remains the perfect match for the metallic Al phase representing the other composite fraction. This metallic Al structural phase fraction can be seen as in the initial stage of nucleation, having very well intercalated into its surface stabilizing GC and crystallized PVP matrix component, respectively. Similarly, the Al reference ICDD PDF-2: 04-0787 CIF file generated XRD pattern plotted as Al_generated, matches well with that of the RTSC/3PVP-2Al composite product Al phase completing crystallographic phase identification step. No other impurity phase corresponding to the initial untreated precursor and other reaction generated unwanted phases are observed, even though the sonocrystallization process progress is achieved at RT. Ice cooled chilled water maintained at 20 °C is circulated all around the sonochemical reaction vessel to dissipate bulk solution heat accumulation during 2 h long continuous mode sonochemical processing.

Among the allotropes of carbon, hexagonal GC crystal form in ABABAB… carbon layers stacking sequence is a fascinating microstructural feature enriching extensive research and development activities [390]–[396]. Significantly, this carbon forms layers one above other in parallel stacking (see fig. 4 13 (a)) which makes GC soft and slippery nature due to contributions of these carbon layers facile expansion along the c-axis. These c/2 stacked carbon layers are the x-ray diffracting entities that produce a pronounced (002) diffraction peak, representing layers spacing. Any changes to this c/2 spacing brought in can easily be tracked by XRD measurement. The inset shown in fig. 4 13 (a) is the XRD patterns of graphite, and one of its c-axis expanded structures, indicates this one to one correspondence of c-axis stretching leading to XRD peak shift to lower angles. It is important to note here that for material under stress-strain investigation, in the elastic region below, yield point stress is proportional to strain [397]. A graphical schematic of the generic physical shape for materials stress-strain curve in both the elastic and plastic regions is plotted in fig. 4 13 (c).

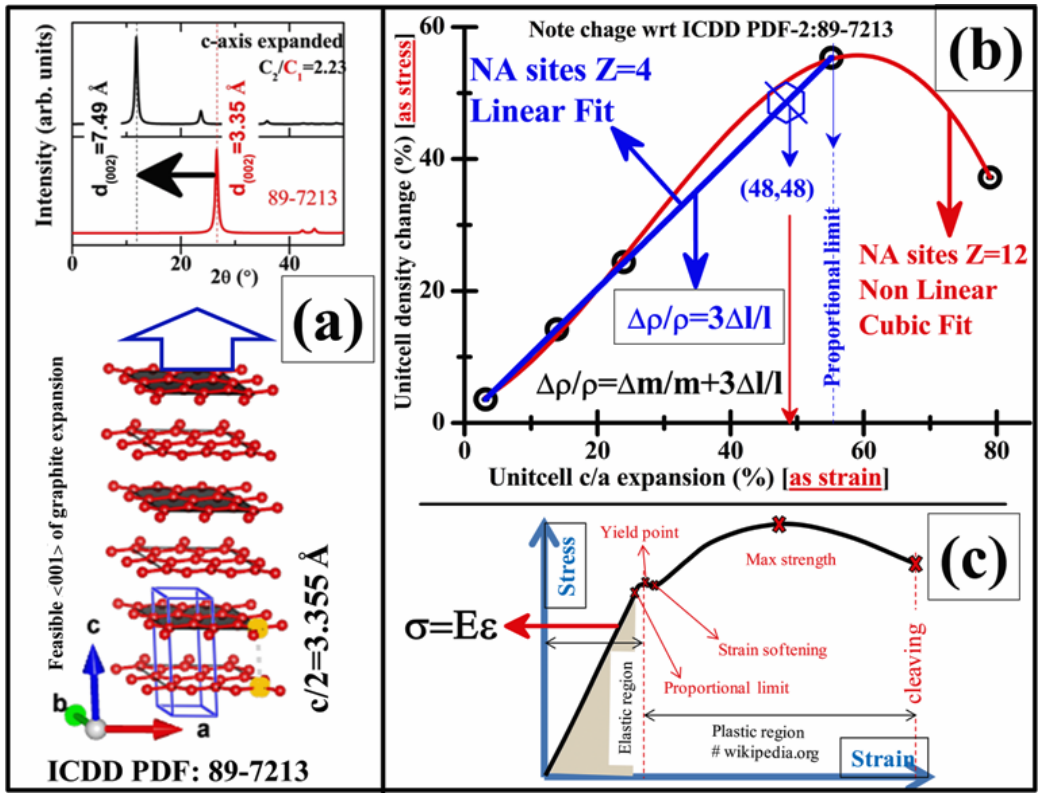

Fig.4 13 [Expandable GC]: (a) schematic of Graphite unit cell and its generated XRD pattern (b) change in unit cell density versus lattice expansion as stress-strain curve, and (c) physical shape of stress-stain curve adopted from Wikipedia respectively.

Thereby in the elastic region, observation the stress versus strain proportionallity behavior observation is analogously extended to the material density changes brought in by c-axis elongation. This presumption is correct until no external mass flows into or out of this hexagonal carbon unit cell (GC, crystal system- hexagonal, space group number-194, space group symbol-P63/mmc) is strictly prohibited. That is, elongation proportionally reduces unit-cell density. Based on this fact, both the c-axis elongation and corresponding possible density changes of a variety of GC unit-cells are plotted in fig. 4 13 (b). The expansions of these GC unit-cells are with respect to the ICDD PDF-2: 89-7213 standard file. The set of expandable GC unit cells utilized are tabulated in Table-1 taken from "The material project: A material genome approach to accelerating material innovation" [398]. Two implications of this correlation are; (1) a nonlinear cubic power law is the best fit (red curve, $\chi^2$ =0.999) to the entire dataset considered, when there is mass flow into the unit cell. That is, once the number of carbon atomic sites in the unit cell is increased from z=4 to z=12. The physical appearance of both the plots of fig. 4 13 (b) and (c) becomes analogous. (2) A linear fit is the best fit (blue line, $\chi^2$ =0.999) untill the GC lattice expansion reaches 55 % (proportional limit); and the number of GC atomic site is maintained at z=4. An unit cell expansion of less than 55 % is recoverable, and the expandable graphite is in the elastic region. In the fabricated RTSC/PVP-Al composite $d_{(002)}$=7.66 Å (see Fig.4.10 GC_ generated) represents 48 % elongation, thereby is in the elastic region. One-to-one correspondence employing fig. 4 13 (b) plots, it is estimated that the expanded GC density must be 1.17 g/cm³. Thereby the material project mp-99182 file represents the ideal current sonocrystallized expanded GC unit cell parameters.

Table 4 1 GC unit cell taken from the material project (mp) [398] and ICSD database.

| **ID mp-48** | **ID mp-606949** | **ID mp-997182** | **ICSD-426931** | **ICSD-617290** | **ICDD-897213** |
|---|---|---|---|---|---|
| a=b=2.467 Å c=7.803 Å | a=b=2.467 Å c=31.983 Å | **a=b=2.468 Å c=14.998 Å** | a=b=2.469 Å c=8.841 Å | a=b=2.470 Å c=6.930 Å | a=b=2.464 Å c=6.711 Å |
| α=β=90° γ=120° | α=β=90° γ=120° | **α=β=90° γ=120°** | α=β=90° γ=120° | α=β=90° γ=120° | α=β=90° γ=120° |
| Z=4 | Z=12 | **Z=4** | Z=4 | Z=4 | Z=4 |
| ρ=1.94 g/cm³ | ρ=1.42 g/cm³ | **ρ=1.01 g/cm³** | ρ=1.71 g/cm³ | ρ=2.18 g/cm³ | ρ=2.26 g/cm³ |

This illustration of the soft and expandability feature of the GC, by bringing a correlation with well-established materials stress-strain plot is most illustrative. This analysis also stands in justification and support of its broad applicability to the field of batteries as an electrode material, where repeated charging and discharging are linked to reversible expansion/ contraction of graphite composite electrodes [399]–[401].

### 4.4.3 Intercalation of metallic Al in sonocrystallized GC and PVP Matrix composite

The sonication generated self-heating (SH) is utilized as one of the effective means to facilitate nanocrystalline Al growth, suitably embedded, and stabilized in the sonocrystallized GC and PVP Matrix fraction delivering required 1:1= polymer(P) to metal(M) composite. A quantifying parameter, i.e., "degree of crystallinity" (DOC) representing only the Al crystalline phase fraction, is evaluated to justify the synthesis of the desired composite. For example, the P: M=1:1, 4:1 composites based on the definition must have DOC of about 50 and 20 % of Al, respectively. The XRD data are shown in fig. 4 14 highlights two distinguishable processing aspects; (1) PVP fraction sonocrystallization at RT, (2) metallic Al crystal growth utilizing the bulk heating generated by the continuous mode 2 hrs sonochemical processing. The Al grown phase fraction reaches DOC= 49 % is as per the desired P: M=1:1 composite.

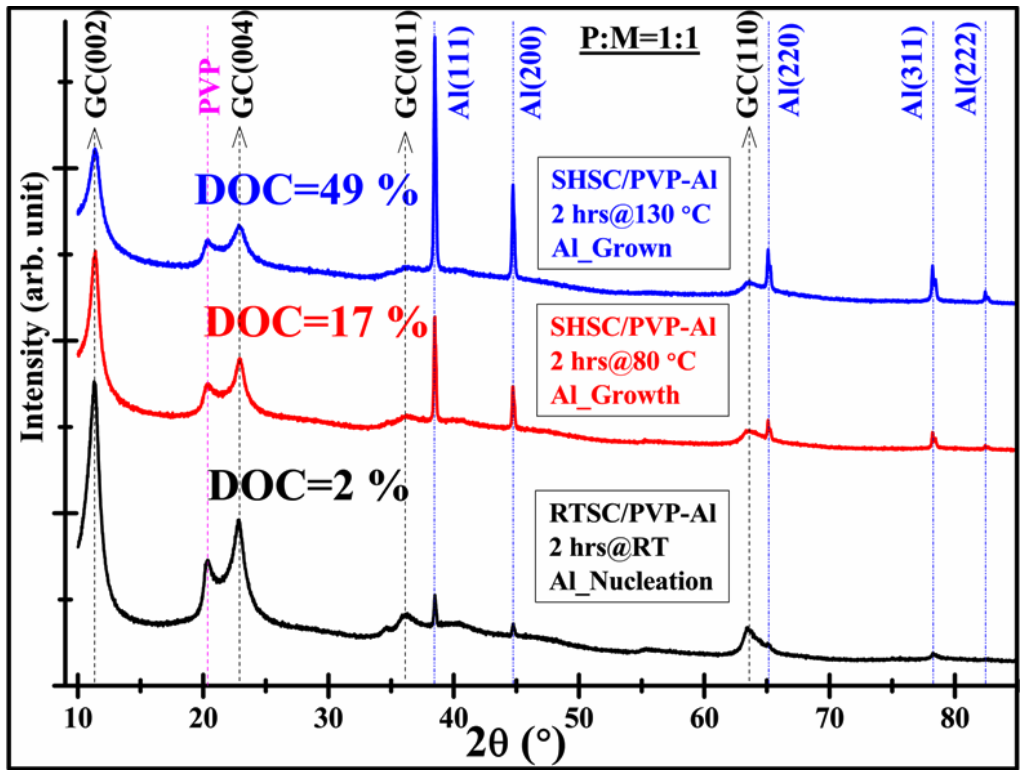

Fig.4 14 [*Intercalation of metallic Al*]: Metallic Al nucleation and growth by 2hrs sonication generated self heating (SH). Composite processed at RT (RTSC/PVP-Al), 80 °C (SHSC/PVP-Al), 130 °C (SHSC/PVP-Al) respectively.

It is pertinent to mention here that two sonochemical SH temperatures 80 °C and 130 °C respectively, reached after 1 h and 2 hrs of processing, are utilized for Al crystal growth. Also, to illustrate DOC values computation, two processed P: M fraction XRD data ($Y_{xrd}$) is shown in fig. 4 15. The CIF of the identified crystallized structural phases of GC, PVP, and Al are used to generate the whole XRD pattern. Each structural phase is refined, and the individual peak phase is generated using fundamental parameters profile fitting (FPPF) approach [374]. The extracted DOC of 56 and 22 % are as per the fraction of 1:1 and 4:1 chosen for P: M, respectively. The obtained final profile fit ($Y_{cal}$), the difference pattern ($Y_{xrd}-Y_{cal}$), and along with goodness of fit index ($\chi^2$) is shown in fig. 4 15 (a), (b).

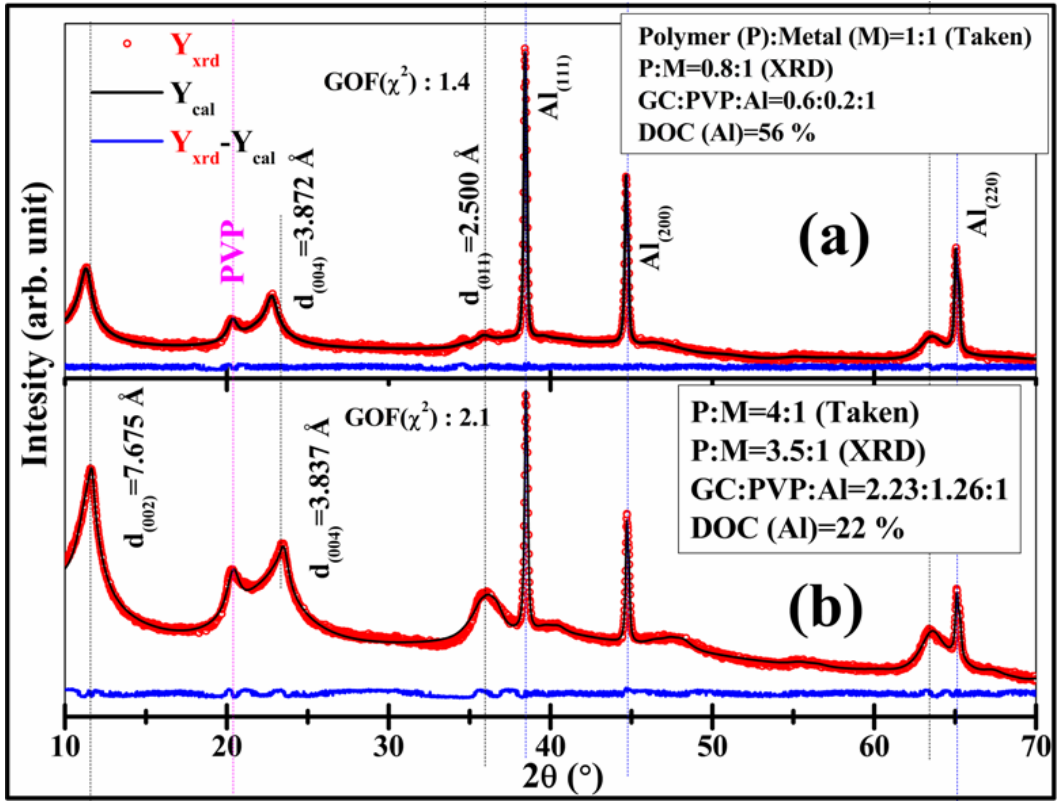

Fig.4 15 [*A set of P: M fraction*]: XRD patterns of the (a) P: M=1:1, (b) P: M=4:1 composites respectively.

## 4.5 Metallic Al Crystal Growth

The synthesized RTSC/PVP-Al composite having the least DOC=2 % of Al, is the precursor chosen to illustrate the Al crystal growth. In fact the RTSC/PVP-Al composite is having the Al phase is at its nucleating state (Al_Nucleation). To facilitate Al crystal growth the sonochemical processing generated solution self heating is considered. The 80 °C reached with 1 h of processing is maintained another 1 h. A total 2 hrs of processing at 80 °C increases the DOC to 17 % representing Al growth (Al_Growth). In contrast 130 °C reached during 2 hrs of processing further increases DOC to 49 %, almost approaching the 50 % theoretical DOC limit chosen. Therefore, the DOC=49 % achieved product is identified as Al_Grown. Clearly these XRD quantitative DOC data extracted from the product XRD patterns shown in fig. 4 14, can be identified with Al nucleation, growth and grown features respectively, in the absence of any crystal growth mechanistics.

In the present context, the feasible way to provide a mechanistic understanding of crystal growth is to employ an appropriate tool that facilitates crystallization. One is the utilization of the TEM electron beam (e-beam) irradiation. There are many reports of localized crystallization under TEM e-beam [402]–[406]. The progress of amorphous to crystalline phase transition under TEM e-beam is divided into two categories. These are; (1) (beam energy is large to overtake displacement energy) the crystallization is achieved by the creation/annihilation of point defects and inducing increased atomic mobility [404], [407]–[409], or (2) (for lower beam energy not sufficient for creating atomic displacements) crystallization gets initiated at the amorphous to crystalline interface with the breaking of incorrectly formed interfacial bonds and subsequently rearranges itself to regular crystalline order [403], [410]–[415]. The reason for athermal nature of this TEM e-beam induced crystallization and also why an amorphous (of high relative internal energy) material ends up into an ordered crystalline structure under continuous e-beam impetus can be found elsewhere [402], [416], [417]. Computed experimental data suggest to create point defects in crystalline Al displacement energy of 19 eV is required corresponding to 210 keV primary TEM e-beam [418]. But in the present case of amorphous material having differing local environment than its crystalline form, the displacement energy can be as low as 10 eV [419]. Therefore having 200 keV TEM e-beam operating at step-4 emission mode with well above the predicted displacement threshold energy is expected to create the required effect. It suggests achieved amorphous to crystalline transition is dominantly controlled by point defects creation and annihilation, thereby falls in category 1, as stated. With this brief TEM e-beam irradiation, an athermal crystallization enhancement (DOC increase) tool appropriate to mimic the actual Al crystal growth observed by sonication generated SH can be simulated.

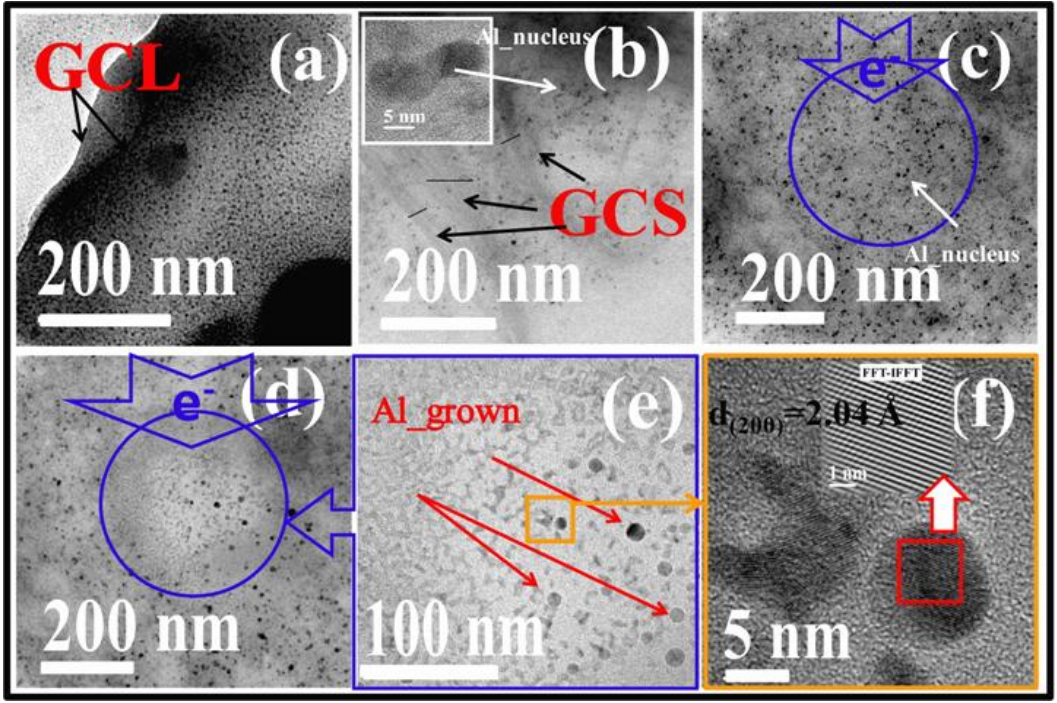

Fig.4 16 [*Nano-Al crystal nuclei*]: synthesized RTSC/PVP-Al composite TEM analysis.

The RTSC/PVP-Al composite having DOC=2 % representing Al is imaged in TEM-BF mode, and the micrographs are shown in figs. 4 16 (a)-(c) respectively. GC in layer (fig. 4 16 (a)) and stacking (fig. 4 16 (b)) having 5-8 nm dark spots well embedded densely packed and uniformly spread can be seen. One of the HR-TEM imaging of these dark spots suggests dense liquid-like material embedding and its flow behavior, having no signature of Al lattice fringes. The inset in fig. 4 16 (b) contains one such Al nucleus (Al_nucleus) in HR-TEM observation. In order to facilitate crystal growth employing TEM e-beam, the protocol schematized by the present author in the previous chapter-III (section3.1.4) is followed [420]. In the TEM BF micrograph shown in fig. 4 14 (c), the blue encircled region is TEM e-beam irradiated (E4I5M) for 5 minutes in HR-TEM mode with step-4 LaB$_6$ electron emission current. The micrograph shown in fig. 4 16 (e) is the e-beam irradiated region from which there is disappearance of black spots (liquid-like containment), undergoes crystallization leading to the growth of spherical Al nanoparticles. The grown spherical Al nanoparticles are of 15-18 nm in diameter. The central section of the E4I5M irradiated region shown in fig. 4 16 (e) is further probed for crystallinity development using HR-TEM mode. The obtained HR-TEM micrograph shown in fig. 4 16 (e), indicates the e-beam irradiation grown Al nanoparticles are crystalline and have lattice fringes of Al d-spacing 2.04 Å. This observation is in concurrence with earlier reports on crystallinity development employing TEM e-beam irradiation as a localized tool.

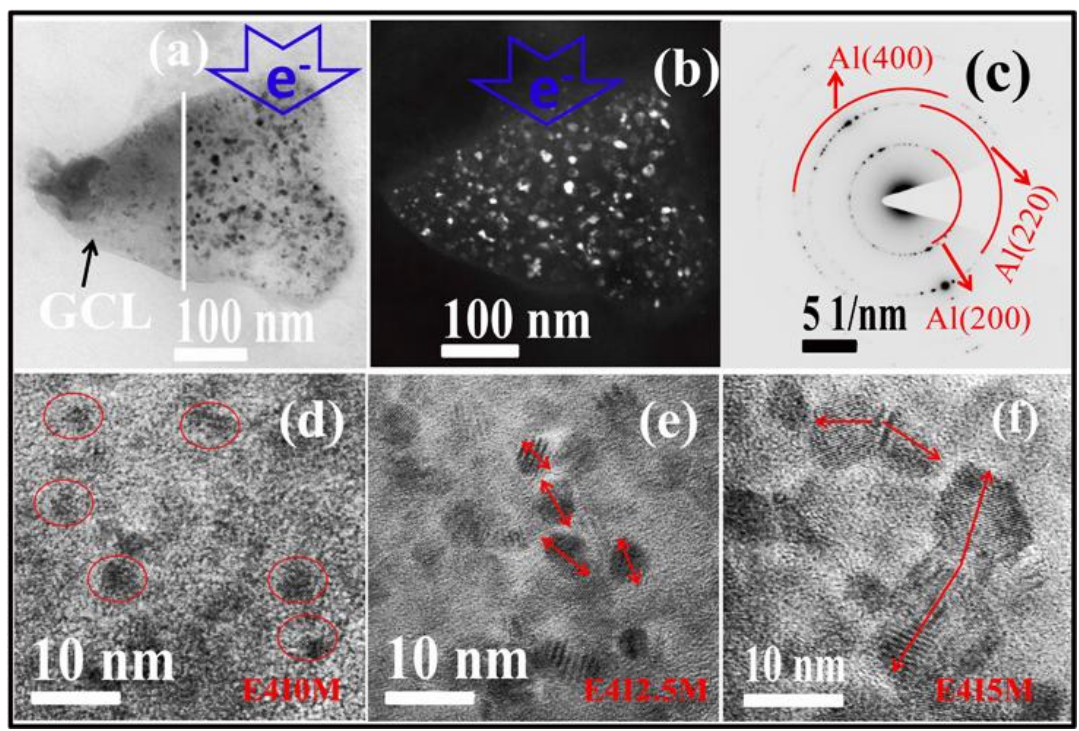

Fig.4 17 [*Al Crystal growth under TEM e-beam*]: synthesized RTSC/PVP-Al composite TEM analysis after exposure to TEM e-beam; (d)-(f) snapshot of the same region illustrating Al crysal growth, (a), (b) demonstrate TEM e-beam gradual movement from right to left facilitating growth in a GCL respectively.

To gain further insight, whether Al crystal growth is by classical Ostwald's ripening (OR) or by particle mediated non-classical (OA) scheme TEM microstructural characterization is employed [420]–[424]. A GC flake having embedded Al nuclei of RTSC/PVP-Al composite shown in fig. 4 17 (a) is half portion (TEM BF) and the portion that is subsequently completely E4I5M e-beam irradiated is shown in fig. 4 17 (b). The observed clear brighter spots in TEM DF imaging mode all over the GC flake validates the crystallinity of embedded nanoparticulate. The entire GC flake portion acquired in TEM SAED mode validates nanoparticulate entities to Al structural phase ring indexing (see fig. 4 17 (c)). Sequential e-beam irradiated RTSC/PVP-Al composite portion after 0, 2.5, and 5 minutes exposure is shown in figs. 4 17 (d)-(f) validates crystal growth and supports particle attachment. TEM e-beam electron transparency in the HR-TEM micrographs of figs. 4 17 (d)-(f) to classify whether the particle attachment is OR or OA scheme. Another GC flake already once E4I5M irradiated having a comparatively larger 10-15 nm size is chosen for crystal growth observation. One of the edge portions of the flake having 9 Al nanocrystallites is shown in fig. 4 18 (b). Subsequent E4I2.5M exposure few smaller ones disappear, highlighting coarsening of smaller ones coarsening by OR scheme. This is further illustrated in a still larger particulate marked as-1 is shown in fig. 4 18 (d). The OR of particles 2, 3, and simultaneous growth and evolution of particle-1 shape is in support of OR, leading to Al crystal growth. This physical evidence demonstrated under TEM e-beam is consistent with literature on metallic particles crystal growth by ultrasonic induced head-on collision facilitated agglomeration, particle fusion by melting, and coalescence [350]–[355]. The similarity being both (TEM e-beam and Ultrasonic) Al crystal growth is by classical OR mechanism. The difference being that the first is athermal, while in the second, localized temperature rise above melting resulting coalescence is the proven reasoning.

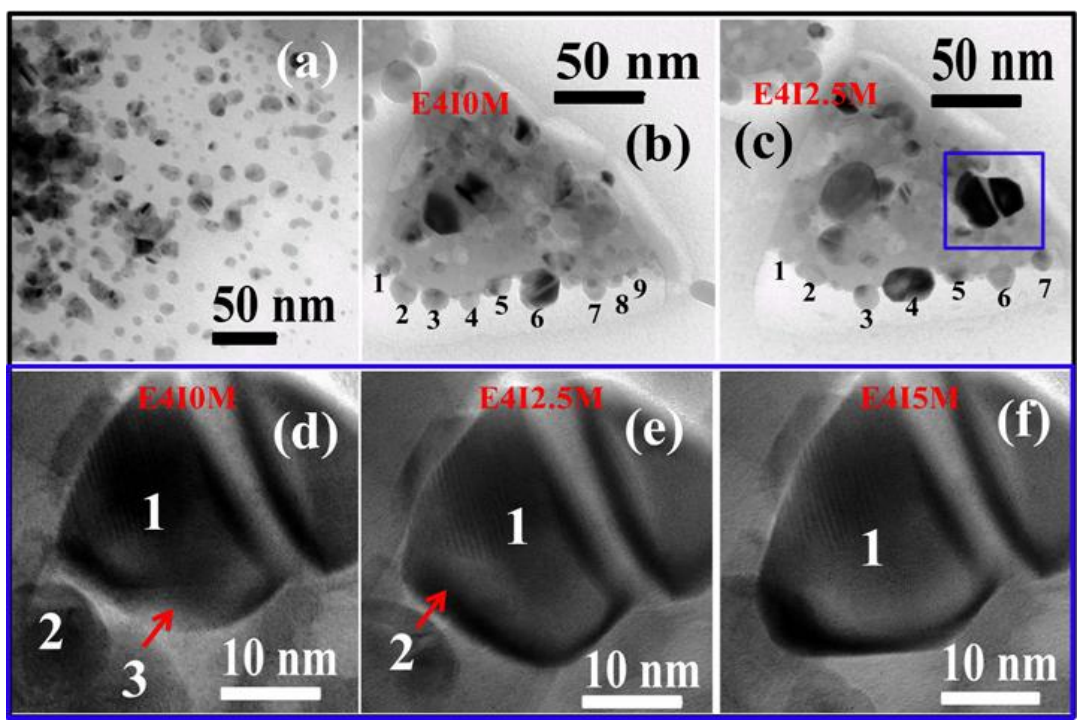

Fig.4 18 [*Nano-Al Crystal growth under TEM e-beam*]: (d)-(f) already exposed larger Al-crystallite is seen to undergoes OR by consuming smaller adjacent ones; (a)-(c) particle attachment illustrations respectively.

## 4.6 Conclusions

The specific conclusions drawn from this chapter in the process of synthesizing air-stable metallic-Al particles embedded in the PVP matrix are listed below.

1) The sonocrystallization of PVP to graphitic carbon (GC) at RT indicates the process as athermal, thereby favors the dominant role of ultrasonic shock waves in causing it.

2) Similarly, the RT processed composite (RTSC/PVP-Al) only has metallic Al in its nucleating state, thereby also in agreement with cited literature that sonocrystallization leads to generation of Al nuclei or a nucleating phase of any sonoprocessed mater.

3) The nano-Al crystal growth is only achieved when the solution is allowed to self-heat during sonoprocessing. The bulk solution heating probably causes an increase in the rate of the head-on collision of these RT generated Al nuclei to fuse. The nuclei fusion generates a crystalline building unit, which subsequently grows by further coalescence based on the duration of sonoprocessing.

4) To validate Al crystals growth by building unit coalescence, Al-rich compositions with Al (M)/ PVP (P) ratio higher than 1:1 ratio investigated indicates building units sono-agglomeration. In this case, the reduced fraction of PVP surfactant offers less hindrance to agglomerate almost 10 nm Al cubes in sidewise fashion to deliver around 359 nm Al 2D-large lumps devoid of an oxide phase. When exposed to TEM e-beam, the de-agglomeration of individual building units is observed.

5) In the case of Al (M)/ PVP (P) fraction= 1:1, the sono-agglomeration of nano-Al building units is actively suppressed by the PVP fraction to deliver approximately 15 nm Al crystallites densely packed inside the PVP matrix. The degree of crystallinity of the Al phase as expected is 56 % (XRD extraction), slightly above the theoretical expected 50 % in line with the composite fraction considered.

6) The arrangement/attachment of nano-Al crystals at the edges of the GC indicates almost all the major features linked to the Al phase like; nucleation, coalescence, and growth mostly happen in the n-hexadecane medium. Simultaneous gradual embedding of grown nano-Al crystals into the GC layers results in intercalation, and leading thereby its c-axis expansion.

7) The crystal structural data of the expandable GC extracted indicates that its expansion is 48 % higher with respect to the standard ICDD structure, to accommodate 56 % nano-Al fraction.

8) The generated composite is air-stable, Al-rich with no amorphous surface oxide and is expected to have many years storability making it suitable for fuel applications.

9) Finally, the conventional protocol-1, which requires around 16 hrs processing time, is brought down to just 2 hrs highlights another demonstration to sonic-assisted process intensification activity.

Quality Parameters," *Food Bioprocess Technol*, vol. 12, no. 5, pp. 839–851, May 2019, doi: 10.1007/s11947-019-02263-5.

[327] A. Chakravorty, "Process intensification by pulsation and vibration in miscible and immiscible two component systems," *Chemical Engineering and Processing - Process Intensification*, vol. 133, pp. 90–105, Nov. 2018, doi: 10.1016/j.cep.2018.09.017.

[328] Z. Wu, S. Tagliapietra, A. Giraudo, K. Martina, and G. Cravotto, "Harnessing cavitational effects for green process intensification," *Ultrasonics Sonochemistry*, vol. 52, pp. 530–546, Apr. 2019, doi: 10.1016/j.ultsonch.2018.12.032.

[329] S. Bhoi, A. Das, J. Kumar, and D. Sarkar, "Sonofragmentation of two-dimensional plate-like crystals: Experiments and Monte Carlo simulations," *Chemical Engineering Science*, vol. 203, pp. 12–27, Aug. 2019, doi: 10.1016/j.ces.2019.03.070.

[330] A. Klaue *et al.*, "Ziegler–Natta catalyst sonofragmentation for controlling size and size distribution of the produced polymer particles," *AIChE J*, vol. 65, no. 9, p. e16676, Sep. 2019, doi: 10.1002/aic.16676.

[331] N. Edwin and P. Wilson, "Investigations on sonofragmentation of hydroxyapatite crystals as a function of strontium incorporation," *Ultrasonics Sonochemistry*, vol. 50, pp. 188–199, Jan. 2019, doi: 10.1016/j.ultsonch.2018.09.018.

[332] X. Dong *et al.*, "Enhanced high-voltage cycling stability of Ni-rich cathode materials via the self-assembly of Mn-rich shells," *J. Mater. Chem. A*, vol. 7, no. 35, pp. 20262–20273, Sep. 2019, doi: 10.1039/C9TA07147D.

[333] H. N. Kim and K. S. Suslick, "The Effects of Ultrasound on Crystals: Sonocrystallization and Sonofragmentation," *Crystals*, vol. 8, no. 7, p. 280, Jul. 2018, doi: 10.3390/cryst8070280.

[334] H. N. Kim and K. S. Suslick, "Sonofragmentation of Ionic Crystals," *Chemistry – A European Journal*, vol. 23, no. 12, pp. 2778–2782, 2017, doi: 10.1002/chem.201605857.

[335] R. Gao, I. Gupta, and E. S. Boyden, "Sonofragmentation of Ultrathin 1D Nanomaterials," *Particle & Particle Systems Characterization*, vol. 34, no. 1, p. 1600339, 2017, doi: 10.1002/ppsc.201600339.

[336] J. Jordens, T. Appermont, B. Gielen, T. Van Gerven, and L. Braeken, "Sonofragmentation: Effect of Ultrasound Frequency and Power on Particle Breakage," *Crystal Growth & Design*, vol. 16, no. 11, pp. 6167–6177, Nov. 2016, doi: 10.1021/acs.cgd.6b00088.

[337] Y. Xi *et al.*, "Sonocrystallization of conjugated polymers with ultrasound fields," *Soft Matter*, vol. 14, no. 24, pp. 4963–4976, Jun. 2018, doi: 10.1039/C8SM00905H.

[338] S. Nalesso, M. J. Bussemaker, R. P. Sear, M. Hodnett, and J. Lee, "A review on possible mechanisms of sonocrystallisation in solution," *Ultrasonics Sonochemistry*, vol. 57, pp. 125–138, Oct. 2019, doi: 10.1016/j.ultsonch.2019.04.020.

[339] J. Jordens *et al.*, "Sonocrystallisation: Observations, theories and guidelines," *Chemical Engineering and Processing - Process Intensification*, vol. 139, pp. 130–154, May 2019, doi: 10.1016/j.cep.2019.03.017.

[340] J. Lee, K. Yasui, M. Ashokkumar, and S. E. Kentish, "Quantification of Cavitation Activity by Sonoluminescence To Study the Sonocrystallization Process under Different Ultrasound Parameters," *Crystal Growth & Design*, vol. 18, no. 9, pp. 5108–5115, Sep. 2018, doi: 10.1021/acs.cgd.8b00547.

[341] C. Backes *et al.*, "Guidelines for Exfoliation, Characterization and Processing of Layered Materials Produced by Liquid Exfoliation," *Chem. Mater.*, vol. 29, no. 1, pp. 243–255, Jan. 2017, doi: 10.1021/acs.chemmater.6b03335.

[342] C. Gibaja *et al.*, "Few-Layer Antimonene by Liquid-Phase Exfoliation," *Angewandte Chemie International Edition*, vol. 55, no. 46, pp. 14345–14349, 2016, doi: 10.1002/anie.201605298.